\documentclass[12pt]{myels}
\usepackage{epsfig}
\newcommand{\nn}{\nonumber}

\newcommand{\be}{\begin{equation}}
\newcommand{\ee}{\end{equation}}
\newcommand{\bea}{\begin{eqnarray}}
\newcommand{\eea}{\end{eqnarray}}
\newcommand{\aprle}{\stackrel{<}{{}_\sim}}

\newcommand{\gev}{ {\rm GeV} }

\begin{document}



\begin{frontmatter}
\begin{center}

\title{Physical Reach of a Neutrino Factory 
in the 2+2 and 3+1 Four-Family Scenario}




\begin{center}
\author[a]{A. \snm Donini}\footnote{E-mail andrea.donini@roma1.infn.it} and
\author[a]{D. \snm Meloni}\footnote{E-mail davide.meloni@roma1.infn.it} 
\address[a]{I.N.F.N., Sezione di Roma I and Dip. Fisica, 
Universit\`a di Roma ``La Sapienza'', P.le A. Moro 2, I-00185, Rome, Italy}
\end{center}



\begin{abstract}
We compare the physical reach of a Neutrino Factory
in the 2+2 and 3+1 four--family models, with similar results in the two schemes;
in both cases huge CP-violating effects can be observed 
with a near detector in the $\nu_\mu \to \nu_\tau$ channel. 
We also study the capability of long baseline experiments
(optimized for the study of the three--family mixing parameter
space) in distinguishing a three (active) neutrino model 
from a four--family scenario.
\end{abstract}


\begin{keyword}
NUFACT01, neutrino, sterile, oscillations, CP-violation.
\end{keyword}

\end{center}
\end{frontmatter}


\setcounter{footnote}{0}

%
%
%
The present experimental solar and atmospheric neutrino data
give strong indications in favour of neutrino oscillations. 
In addition, the results of the LSND experiment \cite{Aguilar:2001ty} 
would imply the existence of a puzzling fourth, sterile, neutrino state. 
There are two very different classes of spectra with four massive neutrinos: 
two pairs of almost degenerate neutrinos divided by the large LSND mass gap 
(the 2+2 scheme); or three almost degenerate neutrinos and an isolated fourth one 
(the 3+1 scheme). The former gave a better fit, as was shown in \cite{Grimus:2001mn}, 
but the recent SNO results \cite{Ahmad:2001an} will certainly restrict the 
allowed parameter region. The latter is at present only marginally compatible 
with the data \cite{3+1citations}. 

An experimental set-up capable of precision measurement of the whole three-neutrino 
mixing parameter space (including the CP violating phase $\delta$)
is under study. This experimental programme consists of the development
of a ``Neutrino Factory'' (high-energy muons decaying in the straight section 
of a storage ring and producing a very pure and intense neutrino beam
\cite{Geer:1998iz,DeRujula:1999hd}) and of suitably optimized detectors. 
We shall consider in what follows a neutrino beam 
resulting from the decay of $n_\mu = 2 \times 10^{20}$ unpolarized 
positive and/or negative muons per year. The collected muons have energy $E_\mu$ 
in the range $10 - 50$ GeV. 

The following parametrizations were adopted for the four--family mixing matrix:
\vspace{-0.5cm}
\bea
U^{(3+1)}_{PMNS} &=& U_{14} (\theta_{14}) \; 
            U_{24} (\theta_{24}) \; 
            U_{34} (\theta_{34}) \; 
\times 
            U_{23} (\theta_{23}\, , \, \delta_3) \; 
            U_{13} (\theta_{13}\, , \, \delta_2) \; 
            U_{12} (\theta_{12}\, , \, \delta_1) \, ,
\label{3+1param} \\
\nn \\
U^{(2+2)}_{PMNS} &=& U_{14} (\theta_{14}) \; 
            U_{13} (\theta_{13}) \; 
            U_{24} (\theta_{24}) \; 
            U_{23} (\theta_{23}\, , \, \delta_3) \; 
\times
            U_{34} (\theta_{34}\, , \, \delta_2) \; 
            U_{12} (\theta_{12}\, , \, \delta_1) \, ,
\label{2+2param}
\eea
with $\nu_1$ the lightest state and $\nu_4$ the heaviest. 

\section*{If LSND is confirmed \dots}

\dots by MiniBooNE \cite{Church:1997jc}, 
a short baseline experiment is mandatory to explore the four--family
mixing parameter space \cite{Donini:1999jc}.
A comparison of the physical reach at a Neutrino Factory 
in the 2+2 and 3+1 scheme has been extensively presented in \cite{Donini:2001xy}. 

To study the CP-conserving four--family mixing matrix, we
consider a short baseline experiment with an hypothetical 1 ton detector with $\tau$ 
tracking and ($\mu, \tau$) charge identification capability, with constant background 
at the level of $10^{-5}$ and a constant efficiency 
$\epsilon_\mu = 0.5$ for $\mu^\pm$ and $\epsilon_\tau = 0.35$ for $\tau^\pm$. 
Our results show that the considered set-up can severely 
constrain the whole four-family model 
CP-conserving parameter space, both in the 2+2 scheme and 3+1 scheme.
The sensitivity reach to all gap-crossing angles in the 
LSND-allowed region is at the level of $\sin^2 \theta \geq 10^{-6} - 10^{-4}$, 
depending on the specific angle considered, for the 2+2 scheme, and
at the level of $\sin^2 \theta \geq 10^{-5} - 10^{-3}$, in the 3+1 scheme.

\begin{figure}[h!]
\begin{center}
\begin{tabular}{cc}
\epsfxsize6.5cm\epsffile{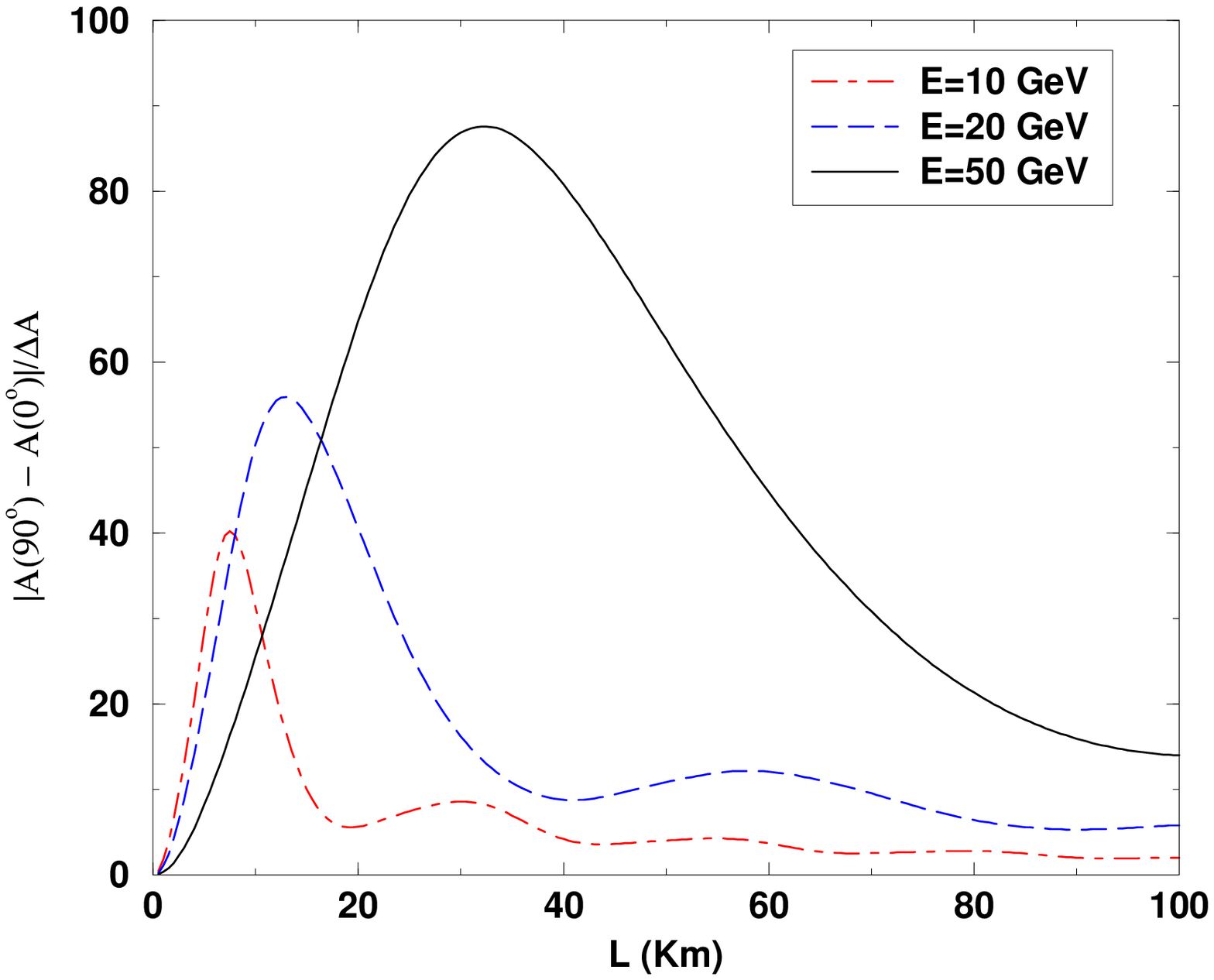} & 
\epsfxsize6.5cm\epsffile{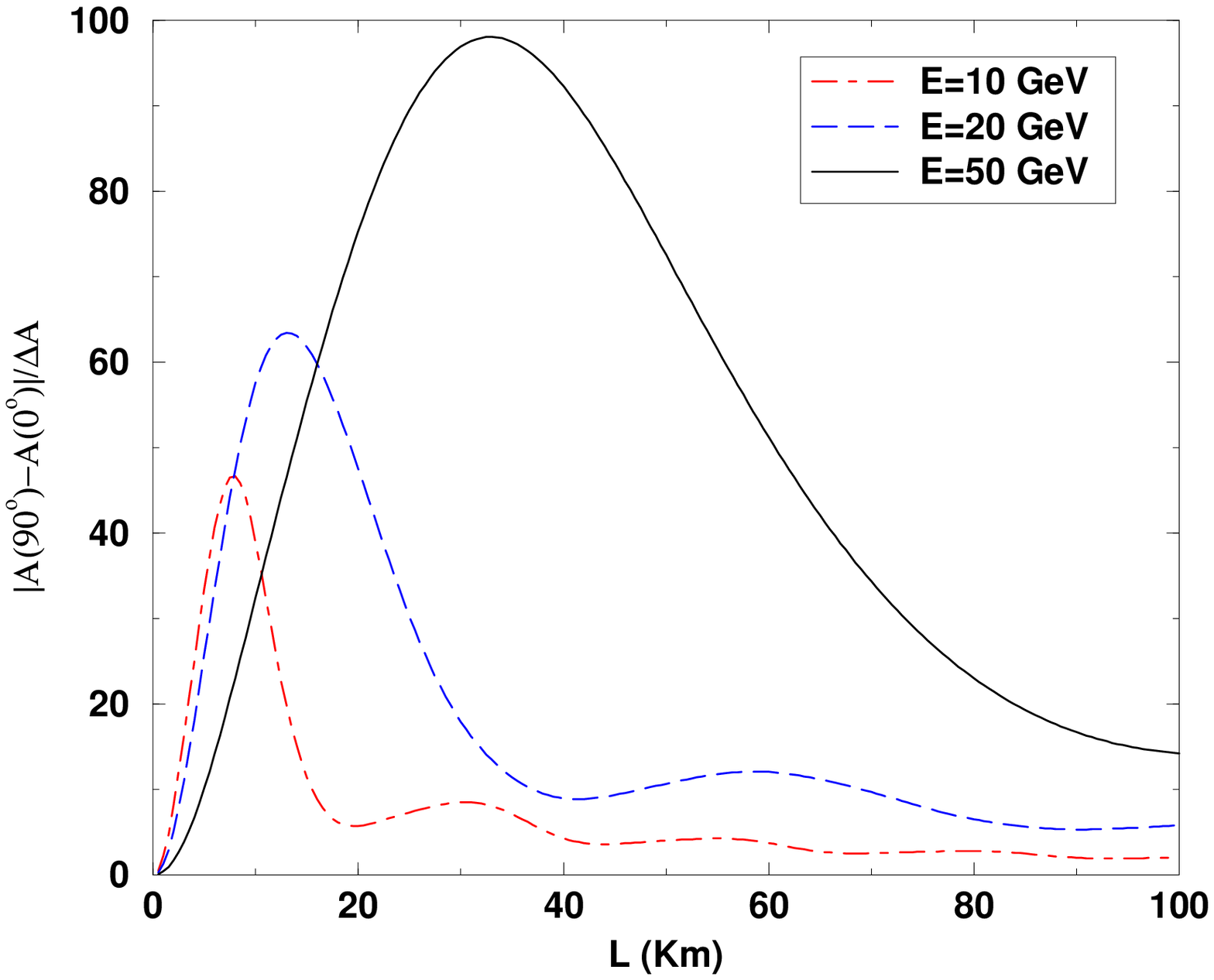}  \\
\end{tabular}
\caption{{\it Signal over statistical uncertainty for maximal
CP violation in the $\nu_\mu \to \nu_\tau$ channel 
in the 2+2 scheme (left) and in the 3+1 scheme (right), 
as a function of the baseline $L$, for three values of the
parent muon energy, $E_\mu = 10, 20$ and $50$ GeV. }}
\label{fig:mutaucp}
\end{center}
\end{figure}

To extend our analysis to the CP-violating four--family parameter space we consider 
an hypothetical 10 Kton detector, located a bit farther 
from the neutrino source, $L = O(10-100)$ Km. 
Large CP violation effects are possible in four--family, since the
overall size of some of the CP-violating observables depends
on $\Delta_{atm}$ and not on $\Delta_\odot$ (as in three families).
In Fig. \ref{fig:mutaucp} we show the signal-to-noise ratio 
of the subtracted integrated CP asymmetry \cite{DeRujula:1999hd} 
in the $\nu_\mu \to \nu_\tau$ channel for the 2+2 (left) and the 3+1 (right) scheme, 
respectively. In both cases, for $E_\mu = 50$ \gev, $\sim 100$ standard deviations 
are attainable at $L \simeq 30 - 40$ Km. 
The other two channels ($\nu_e \to \nu_\mu, \nu_\tau$) give a much smaller significance 
in both schemes. 

\section*{In the absence of a conclusive confirmation of the LSND results \dots}

\dots the detectors will be located very far away from the neutrino source. 
It is of interest to explore the capability of such an experimental set-up to discriminate 
between a three-neutrino model and a possible four--neutrino scenario,
and to understand if three--family data implying a large CP-violating phase
can be also fitted in a CP-conserving four--family model. A detailed 
answer to both questions has been given in \cite{Donini:2001xp}.
The four--family 3+1 scheme reduces to the three--family model
for vanishing mixing with the isolated state, see eq. (\ref{3+1param}), and
a discrimination will be possible only for mixing large enough. 
On the other hand, in the limit of vanishing gap--crossing angles 
the 2+2 scheme mixing matrix, eq. (\ref{2+2param}), reduces to that of 
two independent two--neutrinos oscillations. The possibility of confusion 
is negligible in this case.

We consider a 40 Kton magnetized iron detector with realistic
backgrounds and efficiencies \cite{Cervera:2000vy}. We fit with a three--neutrino 
model ``data'' obtained smearing the four--family theoretical input, following 
ref. \cite{goldenetal}.
In the 3+1 scheme we fixed $\Delta m^{2}_{LSND}=1\;{\rm eV}^2$, 
$\Delta m^{2}_{atm}=2.8\;10^{-3}\;{\rm eV}^2$, $\Delta m^{2}_{\odot}=1\;10^{-4}\;
{\rm eV}^2$,  $\theta_{12}$ = $22.5^{\circ}$ and $\theta_{23}$ = $45^{\circ}$;
we also assumed $\theta_{14} = \theta_{24} = 2^{\circ}, 5^{\circ}, 10^{\circ}$, and 
allowed a variation of $\theta_{13} \in [ 1^\circ, 10^\circ]$ and
$\theta_{34} \in [ 1^\circ, 50^\circ]$. The matter effects have been evaluated with
constant density $\rho = 2.8\,(3.8)\;{\rm g\,cm^{-3}}$ for $L=$732 
or 3500 (7332) Km. The CP violating phases in the 3+1 model have been set to zero.
In the three--neutrino model, we fixed $\theta_{12} = 22.5^{\circ}$ and 
$\theta_{23} = 45^{\circ}$, allowing the remaining parameters to vary in the 
intervals $\theta_{13} \in [1^\circ, 10^\circ], \delta \in [-180^\circ, 180^\circ]$.

\begin{figure}[h!]
\begin{center}
\begin{tabular}{c}
\epsfxsize10cm\epsffile{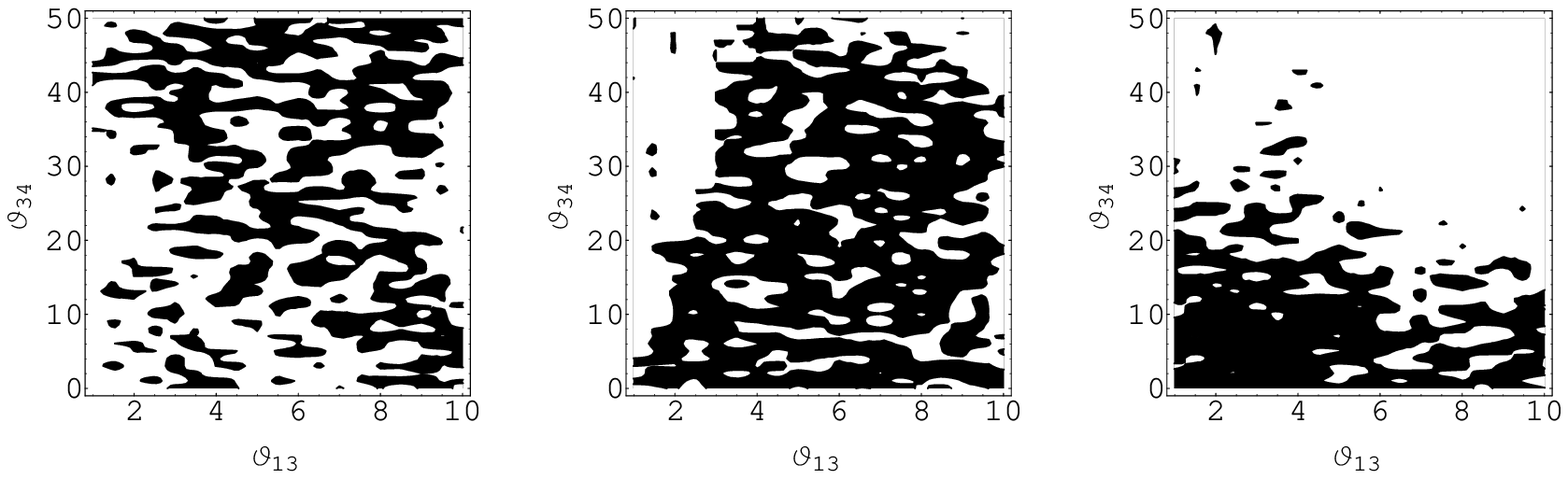} \vspace{-0.5cm} \\
\epsfxsize10cm\epsffile{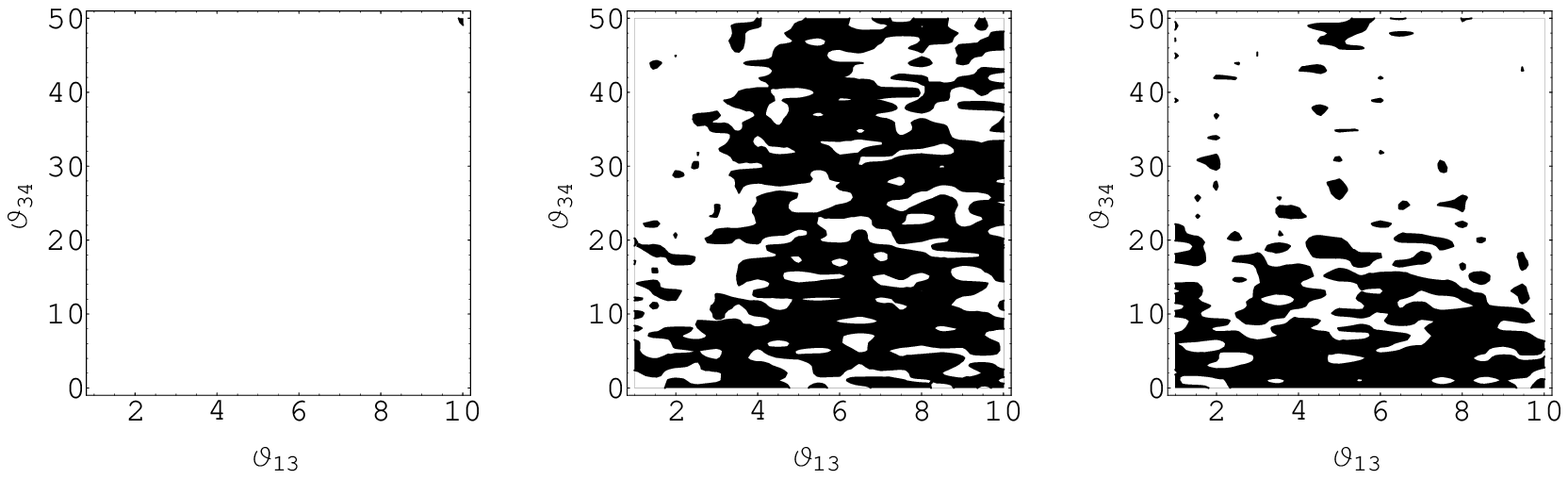} \vspace{-0.5cm} \\
\hspace{-1cm} \epsfxsize9cm\epsffile{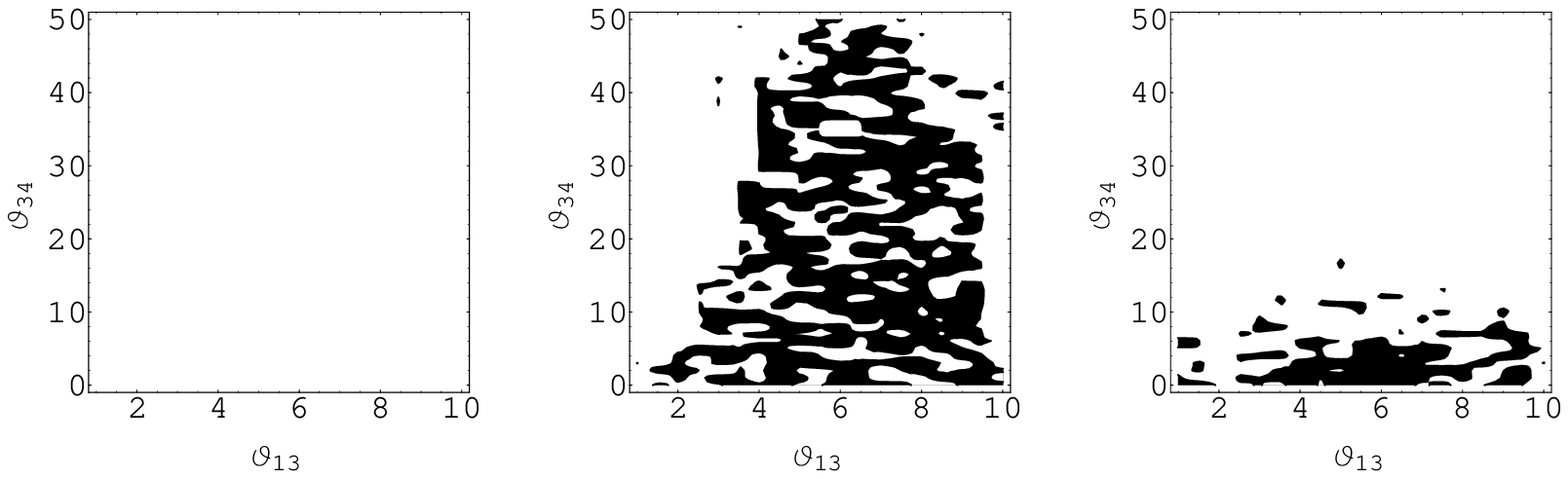} \vspace{0.5cm} \\
\end{tabular}
\caption{{\it ``Dalmatian dog hair'' plots for: 
a) $\theta_{14} = \theta_{24} = 2^\circ$, five bins; 
b) $\theta_{14} = \theta_{24} = 5^\circ$, five bins; 
c) $\theta_{14} = \theta_{24} = 5^\circ$, ten bins; 
from left to right: $L =$ 732, 3500 and 7332 Km.}}
\label{fig:dalmata}
\end{center}
\end{figure}

The results depend heavily on the values of the small gap-crossing 
angles $\theta_{14}$ and $\theta_{24}$. If their value is small ($2^{\circ}$),  
confusion is possible almost everywhere in the $(\theta_{13}, \theta_{34})$ plane. 
This is shown in the first row of Fig.~\ref{fig:dalmata}, 
where in the dark regions of the ``dalmatian dog hair'' plot \cite{disney} 
the three--neutrino model is able to fit at 68\%  c.l. ``data'' generated 
in four families (five energy bins and five years of data taking have been considered).
The reason of the blotted behaviour stands in statistical fluctuations 
in the smearing of the input distributions.
For increasing values of $( \theta_{14}, \, \theta_{24} )$, 
the extension of blotted regions decreases. 
This is shown for $\theta_{14}=\theta_{24}= 5^{\circ}$ in the second row of
Fig.~\ref{fig:dalmata}, again for five energy bins. 
Notice that with the shortest baseline we can always tell 3 from 3+1. 
In the last row of Fig.~\ref{fig:dalmata} we present dalmatian plots
for ``data'' consisting of ten energy bins. The regions 
of confusion are considerably reduced, since the different energy 
dependence in matter helps in the distinction of the two models: 
a fit of the largest baseline data is successful for $\theta_{34} \aprle 10^{\circ}$ only.
A further increase of the gap--crossing angles to $\theta_{14} = 
\theta_{24} = 10^{\circ}$ makes the distinction (even at 95\% c.l.) of the 
two models possible in the whole $(\theta_{13}, \theta_{34})$ plane.

In the blotted regions of Fig. \ref{fig:dalmata}, the corresponding 
fitted value of the CP violating phase $\delta$ is generally not large. 
This is particularly true for $L = 3500$ Km, whereas for $L = 7332$ Km the 
determination of $\delta$ is somewhat looser. 
For the shortest baseline, $L = 732$ Km, we have the largest spread in the 
values of $\delta$, although the most probable value is still close to zero. 
Finally, data that can be fitted with a CP phase close to 90$^\circ$ in the 
three--neutrino theory cannot be described  
in a CP conserving 3+1 theory, provided that   
data at two different distances are used.  



\end{document}